\documentstyle[aps,preprint]{revtex}

\begin{document}

\title{Non-Gaussian Effects on Domain Growth}

\author{Sang Pyo Kim,$^{1,2,}$\footnote{E-mail:
spkim@phys.ualberta.ca; sangkim@ks.kunsan.ac.kr} and F. C. Khanna,$^{1,3,}$\footnote{E-mail:
khanna@phys.ualberta.ca}}

\address{${}^{1}$Theoretical Physics Institute, Department of
Physics, University of Alberta, Edmonton, Alberta, Canada T6G
2J1\\ ${}^{2}$Department of Physics, Kunsan National University,
Kunsan 573-701, Korea \\
${}^{3}$TRIUMF, 4004 Wesbrook Mall, Vancouver, British Columbia,
Canada, V6T 2A3}

\date{\today}

\tightenlines

\maketitle

\begin{abstract}
The vacuum two-point correlation function is calculated beyond the
Gaussian approximation during the second order phase transition.
It is found that the correlation function is dominated by the
Gaussian term immediately after the phase transition but later is
taken over by non-Gaussian terms as the spinodal instability
continues. The non-Gaussian effects lead to larger size domains
and may imply a smaller density of topological defects than that
predicted by the Hartree-Fock approximation.
\end{abstract}


The phenomenon of continuous
phase transitions is common to many diverse areas of physics. In
particular phase transitions have been studied extensively in
condensed matter. However the dynamical process of the onset of
phase transition and temperature dependence or more generally its
time-dependence is less well-understood. Topological defects may be formed during spontaneous symmetry breaking
phase transitions. A theoretical
understanding for defect formation in such phase transitions
was provided by Kibble \cite{kibble} and Zurek \cite{zurek}
on the basis of causality principle.

Since then, it is argued that the formation of vortices may be
observed experimentally in liquid ${}^4 {\rm He}$ \cite{hendry}
and ${}^3 {\rm He}$ \cite{bauerle}. It is well known that liquid
${}^4 {\rm He}$ has a $\lambda$-point phase transition to a
condensed phase, superfluid ${}^4 {\rm He}$, at T = 2.17 K. For
liquid ${}^3 {\rm He}$ the phase transition to a superfluid state
occurs at 2 mK. The transition in ${}^3 {\rm He}$ is more like the
case of superconductivity in metals where pairs of electrons are
coupled to form loosely bound bosons in ${}^1 S$ state, except
that now ${}^3 {\rm He}$ is a fermion and the pairs are in ${}^3
P$ state. However it appears that experiments do not give a
definite result on the vortex formation.

The early Universe might have undergone a sequence of spontaneous symmetry breaking
phase transitions \cite{kibble,vilenkin}. Heavy-ion
collisions at relativistic energies lead to a quark-gluon plasma,
which is believed to mimick the early stage of the Universe.
In the process of cooling this plasma, the system will go
through a chiral phase transition. It is likely that the phase
transition may lead to the formation of disoriented chiral
condensate (DCC) \cite{wilczek}.
The evidence of DCC would be emission
of large number of coherent pions from the decay of such condensates.

Current methods to study the dynamical formation of defects from
the onset of phase transitions are classified largely into the
time-dependent Landau-Ginzburg (TDLG) equation \cite{zurek2} and
quantum field theory (QFT)
\cite{boyanovsky,rivers,bowick,calzetta,kim-lee}. In the former
the dynamics of phase transitions is described by the classical
TDLG equation with a stochastic noise, whereas in the latter the
dynamics is described by the quantum field equation with or
without the noise. QFT is more fundamental than TDLG equation in
the sense that QFT is the best tested, microscopic theory of
nature. But QFT has been limited to, more or less, the
Hartree-Fock or Gaussian approximation due to technical
difficulty. However, during the second order phase transition via
a quench, the soft (long wavelength) modes grow exponentially. So
it is likely that nonlinear terms would become comparable to the
Gaussian contribution during the phase transition and spinodal
instability quickly turns the dynamics of the problem to be highly
nonlinear and nonperturbative. It may then be anticipated that
large non-Gaussian fluctuations will drive the domain size.

In this Letter we find the nonequilibrium quantum evolution beyond
the Gaussian approximation during the second order phase
transition and calculate the correlation length from the two-point
function. As a QFT model we shall focus on a real scalar field
undergoing an instantaneous quench that describes the dynamical
formation of domains.  For that purpose we use the Fock basis
obtained in Refs. \cite{kim-lee,kim,kim-lee2} to find
perturbatively and systematically the improved vacuum state during
the quench of the second order phase transition. Our scheme is
more systematic than an improved Gaussian state corrected by the
second excited state \cite{cheetham}, since we can find all the
corrections to the Gaussian state to any desired order of the
coupling constant using the standard perturbation method.

The QFT model for the second order phase transition is described
by the Hamiltonian
\begin{equation}
H(t) = \int d^3{\bf x} \Biggl[\frac{1}{2}\pi^2 + \frac{1}{2}
(\nabla \phi)^2 + \frac{m^2 (t)}{2} \phi^2 + \frac{\lambda}{4!}
\phi^4 \Biggr], \label{ham}
\end{equation}
where
\begin{equation}
 m^2 (t) = \cases{ m_i^2, & $ t < 0$, \cr -
m_f^2, & $t
> 0$. \cr}
\end{equation}
The quantum dynamics of nonequilibrium process is prescribed by
the time-dependent functional Schr\"{o}dinger equation
\begin{equation}
i \hbar \frac{\partial}{\partial t} \vert \Psi (t) \rangle =
\hat{H} (t) \vert \Psi(t) \rangle. \label{sch eq}
\end{equation}
By redefining the Hermitian Fourier modes as
\begin{equation}
\phi_{\bf k}^{(+)} (t) = \frac{1}{2} [\phi_{\bf k} (t) + \phi_{-
{\bf k}} (t)], \quad \phi_{\bf k}^{(-)} (t) =
\frac{i}{2}[\phi_{\bf k} (t) - \phi_{- {\bf k}} (t)],
\end{equation}
where
\begin{equation}
 \phi ({\bf x}, t) = \int \frac{d^3 {\bf k}}{(2 \pi)^{3/2}}
\phi_{\bf k} (t) e^{i {\bf k} \cdot {\bf x}},
\end{equation}
we rewrite the Hamiltonian (\ref{ham}) as
\begin{equation}
H (t) = \sum_{\alpha} \frac{1}{2}\pi_{\alpha}^2 + \frac{1}{2}
\omega_{\alpha}^2 (t) \phi_{\alpha}^2 + \frac{\lambda}{4!}
\Biggl[\sum_{\alpha} \phi_{\alpha}^4 + 3 \sum_{\alpha \neq \beta}
\phi_{\alpha}^2 \phi_{\beta}^2 \Biggr], \label{mod ham}
\end{equation}
where
\begin{equation}
\omega_{\alpha}^2
(t) = m^2 (t) + {\bf k}^2,
\end{equation}
with $\alpha$ and $\beta$ denoting $\{{\bf k}, (\pm) \}$. In
deriving Eq. (\ref{mod ham}) we neglect the odd power terms from
the self-interaction whose expectation values vanish with respect
to symmetric states because we are interested in improving the
Gaussian state that preserves the parity. The mode-decomposed
Hamiltonian (\ref{mod ham}) consists of coupled quartic
oscillators, the coupling among different modes arising from the
last term. Because of the time-dependent mass squared  the
Hamiltonian (\ref{mod ham}) describes a system in nonequilibrium,
the degree of nonequilibrium being determined by the rate of
change of the mass. In order to handle this nonequilibrium system
we use a recently introduced canonical method, the so-called
Liouville-von Neumann approach \cite{kim-lee,kim,kim-lee2}.

The essential idea of Refs. \cite{kim-lee,kim,kim-lee2} is to
introduce the following time-dependent annihilation and creation
operators for each mode that satisfy the quantum Liouville-von
Neumann equation:
\begin{equation}
\hat{A}_{\alpha} (t) = \frac{i}{\sqrt{\hbar}} [\varphi^*_{\alpha}
(t) \hat{\pi}_{\alpha} - \dot{\varphi}_{\alpha}^* (t)
\hat{\phi}_{\alpha} ], \quad \hat{A}^{\dagger}_{\alpha} (t) = -
\frac{i}{\sqrt{\hbar}} [\varphi_{\alpha} (t) \hat{\pi}_{\alpha} -
\dot{\varphi}_{\alpha} (t) \hat{\phi}_{\alpha} ].
\end{equation}
Then the Hamiltonian (\ref{mod ham}) can be represented in terms
of $\hat{A}_{\alpha}, \hat{A}_{\alpha}^{\dagger}$ and divided into
\begin{equation}
\hat{H} (t) = \hat{H}_{\rm G} (t) + (\lambda \hbar^2) \hat{H}_{\rm
P} (t),
\end{equation}
where $\hat{H}_{\rm G}$ is the quadratic (Gaussian) part and
$\hat{H}_{\rm P}$ is the quartic part:
\begin{eqnarray}
\hat{H}_{\rm  G} &=& \frac{\hbar}{2} \sum_{\alpha}
\Biggl[(\dot{\varphi}_{\alpha}^{2} + \omega^2_{\alpha}
\varphi_{\alpha}^{2})\hat{A}^{2}_{\alpha} + 2
(\dot{\varphi}_{\alpha}^* \dot{\varphi}_{\alpha} +
\omega_{\alpha}^2 \varphi_{\alpha}^* \varphi_{\alpha})
\hat{A}^{\dagger}_{\alpha} \hat{A}_{\alpha} +
(\dot{\varphi}^{*2}_{\alpha} + \omega^2_{\alpha}
\varphi^{*2}_{\alpha})\hat{A}^{\dagger 2}_{\alpha} \Biggr]
\nonumber\\ &&+ \frac{\lambda \hbar^2}{4} \Bigl(\sum_{\beta}
\varphi_{\beta}^* \varphi_{\beta} \Bigr) \sum_{\alpha}
\Biggl[\varphi_{\alpha}^2 \hat{A}^{2}_{\alpha} + 2
\varphi_{\alpha}^* \varphi_{\alpha}
\hat{A}^{\dagger}_{\alpha}\hat{A}_{\alpha} + \varphi_{\alpha}^{*2}
\hat{A}^{\dagger 2}_{\alpha} \Biggr],
\end{eqnarray}
and
\begin{eqnarray}
\hat{H}_{\rm  P} &=& \frac{1}{4!} \Biggl\{ \sum_{\alpha} \sum_{k =
0}^{4} {4 \choose k} \varphi_{\alpha}^{*(4-k)} \varphi_{\alpha}^k
\hat{A}^{\dagger (4-k) }_{\alpha} \hat{A}^k_{\alpha} \nonumber\\
&&+ 3 \sum_{\alpha \neq \beta} \Bigl(\varphi_{\alpha}^2
\hat{A}^{2}_{\alpha} + 2 \varphi_{\alpha}^* \varphi_{\alpha}
\hat{A}^{\dagger}_{\alpha}\hat{A}_{\alpha} + \varphi_{\alpha}^{*2}
\hat{A}^{\dagger 2}_{\alpha}  \Bigr) \Bigl(\varphi_{\beta}^2
\hat{A}^{2}_{\beta} + 2 \varphi_{\beta}^* \varphi_{\beta}
\hat{A}^{\dagger}_{\beta}\hat{A}_{\beta} + \varphi_{\beta}^{*2}
\hat{A}^{\dagger 2}_{\beta}  \Bigr) \Biggr\}. 
\end{eqnarray}
Here we neglect purely $c$-number terms. The requirement that each
$\hat{A}_{\alpha}$ and $\hat{A}^{\dagger}_{\alpha}$ satisfy the
quantum Liouville-von Neumann equation
\begin{eqnarray}
i \hbar \frac{\partial}{\partial t} 
\hat{A}_{\alpha} (t) + [ \hat{A}_{\alpha} (t) , \hat{H}_{\rm G} (t)] = 0, \nonumber\\ 
i \hbar \frac{\partial}{\partial t} \hat{A}^{\dagger}_{\alpha} (t) + [\hat{A}^{\dagger}_{\alpha} (t) , \hat{H}_{\rm G} (t)] = 0, \label{ln eq}
\end{eqnarray}
leads to the mean field equation for each mode
\begin{equation}
\ddot{\varphi}_{\alpha} (t) + \omega_{\alpha}^2 (t)
\varphi_{\alpha} (t) + \frac{\lambda \hbar}{2} \Bigl( \sum_{\beta}
\varphi_{\beta}^* (t) \varphi_{\beta} (t) \Bigr) \varphi_{\alpha}
(t) = 0. \label{mean eq}
\end{equation}
Further the standard commutation relations $[ \hat{A}_{\alpha}
(t), \hat{A}^{\dagger}_{\beta} (t)] = \delta_{\alpha, \beta}$ are
guaranteed for all times by the Wronskian conditions 
\begin{equation}
\dot{\varphi}_{\alpha}^* (t) \varphi_{\alpha} (t) -
\dot{\varphi}_{\alpha} (t) \varphi_{\alpha}^* (t) = i.
\end{equation}
Then the Fock space for each mode consists of the Gaussian vacuum and
excited states with quantum number $n_{\alpha}$
\begin{equation}
\hat{A}_{\alpha} (t) \vert 0_{\alpha}, t \rangle_{\rm G} = 0,
\quad \vert n_{\alpha}, t \rangle_{\rm G} =
\frac{\hat{A}_{\alpha}^{\dagger
n_{\alpha}}(t)}{\sqrt{n_{\alpha}!}}  \vert 0_{\alpha}, t
\rangle_{\rm G}. 
\end{equation}
The Fock space basis satisfies the orthonormality condition
${}_{\rm G}\langle m_{\alpha}, t \vert n_{\beta}, t \rangle_{\rm
G} = \delta_{\alpha \beta} \delta_{m n}$. Up to a time-dependent
phase factor each state $\vert n_{\alpha}, t \rangle_{\rm G}$
satisfies the Schr\"{o}dinger equation for the corresponding mode
of $\hat{H}_{\rm G}$. The wave functional for $\hat{H}_{\rm G}$ is
a product of such a state for each mode. For instance, the exact
vacuum state $\vert 0, t \rangle_{\rm G} = \prod_{\alpha} \vert
0_{\alpha}, t \rangle_{\rm G}$ for $\hat{H}_{\rm G} (t)$ is the
Gaussian vacuum for the total Hamiltonian $\hat{H} (t)$ in the
Hartree-Fock approximation. Other excited states of the Gaussian
vacuum are the multiparticle states of modes, which are compactly
denoted as
\begin{equation}
\vert \{{\cal N}\}, t \rangle_{\rm G} = \prod_{\{ {\cal N}\}}
\frac{\hat{A}^{\dagger \{{\cal N}\}}(t)}{\sqrt{\{\cal N}\}!} \vert
0, t \rangle_{\rm G}
\end{equation}
where $\{{\cal N}\} = (n_1, \cdots, n_{\alpha}, \cdots)$ with all $n_{\alpha}$ being nonnegative integers, $\hat{A}^{\dagger \{{\cal N}\}} = \hat{A}_1^{\dagger n_1}
\cdots \hat{A}_{\alpha}^{\dagger n_\alpha} \cdots$, and $\{{\cal N}\}!
= \prod_{\alpha} n_{\alpha}!$.

Quantum states beyond the Gaussian approximation can be obtained
by treating $\hat{H}_{\rm P}$ as a perturbation to $\hat{H}_{\rm
G}$. The perturbation $\hat{H}_{\rm P}$ excites and de-excites in
pairs the multiparticle  states, so we look for an exact quantum
state of the form
\begin{equation}
\vert \{{\cal N}\}, t \rangle = \hat{U}
[\hat{A}^{\dagger}_{\alpha} (t), \hat{A}_{\alpha} (t); t, \lambda]
\vert \{{\cal N}\}, t \rangle_{\rm G}. 
\end{equation}
Then the Schr\"{o}dinger equation (\ref{sch eq}) takes the form
\begin{equation}
\Biggl[i \hbar \frac{\partial}{\partial t} \hat{U} (t, \lambda) +
[\hat{U} (t, \lambda), \hat{H}_{\rm G} (t)] - \lambda \hat{H}_{\rm
P} (t) \Biggr] \prod_{\alpha} \vert \{{\cal N}\}, t \rangle_{\rm
G} = 0. 
\end{equation}
Using Eq. (\ref{ln eq}) and technically assuming that the time
derivative does not act on $\hat{A}_{\alpha}^{\dagger} (t)$ and
$\hat{A}_{\alpha}(t)$, the operator $\hat{U}$ satisfies an
interaction picture-like equation
\begin{equation}
i \hbar \frac{\partial}{\partial t} \hat{U} (t, \lambda) =
(\lambda \hbar^2) \hat{H}_{\rm P} (t) \hat{U} (t, \lambda).
\label{int eq}
\end{equation}
The formal solution to Eq. (\ref{int eq}) can be written as
\begin{equation}
\hat{U} (t, \lambda) = {\rm T} \exp\Biggl[- \lambda \hbar \int
\hat{H}_{\rm P} (t) dt \Biggr], \label{form sol}
\end{equation}
where T denotes a time-ordered integral and
$\hat{A}_{\alpha}^{\dagger}(t)$ and $\hat{A}_{\alpha}(t)$ are
treated as if they are constant operators. The operator $\hat{U}$
is a unitary operator, which is a consequence of the Hermitian
$\hat{H}_{\rm p}$ and is easily shown from the formal solution
(\ref{form sol}). The unitarity of $\hat{U}$ preserves the
orthonormality of $\langle \{{\cal M} \}, t \vert \{{\cal N}\}, t
\rangle = \delta_{\{{\cal M}\},\{{\cal N}\}}$. The vacuum state is
then given by
\begin{equation}
\vert 0, t \rangle = \hat{U} (t, \lambda) \vert 0, t \rangle_{\rm
G} \equiv  \sum_{\{{\cal N}\}} U_{\{{\cal N}\}} (t)
 \vert \{{\cal N} \}, t \rangle_{\rm G}. \label{vac st}
\end{equation}
Our vacuum state (\ref{vac st}) is non-Gaussian in that it is a
superposition of the Gaussian vacuum and its excited states
\cite{cheetham}, and further has kurtosis (higher moments)
different from the Gaussian vacuum \cite{lesgourgues}.

Before the quench $(t < 0)$, the Gaussian vacuum for each mode is
given by the solution 
\begin{equation}
\varphi_{i, {\bf k}} (t) = \frac{1}{\sqrt{2 \Omega_{i, {\bf k}}}} e^{-i
\Omega_{i, {\bf k}} t}, 
\end{equation}
where Eq. (\ref{mean eq}) leads to the gap equation and yields the correct
Gaussian vacuum energy \cite{chang}. The contribution of
non-Gaussian terms to the correlation function is of the order of
$ (\lambda \hbar)^2 /(2^{11} \Omega_{i, {\bf k}}^4)$. Thus not
only for the weak coupling limit $(\lambda \ll m_i^2)$ but also
for the strong coupling limit $\lambda \simeq m_i^2$, the
non-Gaussian contribution to the two-point function is quite
negligible. This justifies the validity of the Gaussian vacuum
before the phase transition via the quench. However, after the
quench $(t > 0)$, the soft modes $({\bf k}^2 < m_f^2)$ begin to
grow exponentially due to $m^2 (t) = - m_f^2$. The exponential
growth continues until it reaches the spinodal line where
$(\lambda \hbar) (\varphi_{f, {\bf k}}^* \varphi_{f, {\bf k}})$
from the self-interaction takes over the unstable $- m_f^2$.
Immediately after the quench we find approximately the soft mode
solutions to Eq. (\ref{mean eq}) as
\begin{equation}
\varphi_{f,{\bf k}} (t) \approx \frac{1}{2 \sqrt{2 \Omega_{i, {\bf
k}}}} \Biggl[\Biggl(1 - i \frac{\Omega_{i, {\bf
k}}}{\tilde{\Omega}_{f, {\bf k}}} \Biggr) e^{ \tilde{\Omega}_{f,
{\bf k}} t} + \Biggl(1 + i \frac{\Omega_{i, {\bf
k}}}{\tilde{\Omega}_{f, {\bf k}}} \Biggr) e^{- \tilde{\Omega}_{f,
{\bf k}} t} \Biggr], \label{long sol}
\end{equation}
where $\tilde{\Omega}_{f, {\bf k}} \approx \sqrt{m_f^2 - {\bf
k}^2}$. These soft mode solutions after the quench continuously
match those before the quench and become exact for the
non-interacting case $(\lambda = 0)$.

However, during spinodal instability the perturbation $(\lambda
\hbar) \hat{H}_{\rm P}$ grows exponentially, becomes comparable to
the self-interaction term of $\hat{H}_{\rm G}$ and will make a
contribution to the correlation function as much as the Gaussian
contribution. Since the soft mode solution (\ref{long sol}), which
dominates over hard (short wavelength) mode solutions, has an
exponentially growing factor overall time-dependent factor,
$\hat{H}_{\rm P} (t)$ approximately commutes each other for two
different times $t' \neq t$. Hence the formal solution (\ref{form
sol}) has a good approximation
\begin{equation}
\hat{U}_{[1]} (t, \lambda) = \exp \Biggl[- i \lambda \hbar \int
\hat{H}_{\rm P} (t) dt \Biggr]. \label{lin approx}
\end{equation}
Moreover, the higher order terms in an exponential form of the
formal solution (see Appendices of Ref. \cite{kim2}) are
suppressed by the expansion parameter $\lambda \hbar$,  which
consolidates the validity of the operator (\ref{lin approx}) as
far as soft modes are concerned. Using Eq. (\ref{lin approx}) and
finding the most dominant contribution to the two-point
correlation function up to order $(\lambda \hbar)^2$, we obtain
\begin{eqnarray}
\langle 0, t \vert \hat{\phi} ({\bf x}, t) \hat{\phi} ({\bf 0}, t)
\vert 0, t \rangle &=&  \sum_{n = 0}^{\infty} \sum_{\alpha}
\frac{(-i \lambda \hbar)^n}{n!} {}_{\rm G} \langle 0, t \vert [
\int \hat{H}_{\rm P} (t) dt, \hat{\phi}^2_{\alpha} (t)]_{(n)}
\vert 0, t \rangle_{\rm G} e^{i {\bf k} \cdot {\bf x}} \nonumber\\
&\approx& \int \frac{d^3 {\bf k}}{(2 \pi)^3} \hbar \varphi_{\bf
k}^* \varphi_{\bf k} e^{i {\bf k} \cdot {\bf x}} + \frac{(\lambda
\hbar)^2 }{2} \int \frac{d^3 {\bf k}}{(2 \pi)^3} \frac{\hbar}{12}
\Biggl[ \int \varphi_{\bf k}^4 dt \Biggl(\varphi_{\bf k}^*
\varphi_{\bf k} \int \varphi_{\bf k}^{*4} dt \nonumber\\ ~~&-& \varphi_{\bf
k}^{*2} \int \varphi_{\bf k}^{*3} \varphi_{\bf k} dt \Biggr) \int \varphi_{\bf k}^{*4} dt \Biggl(\varphi_{\bf
k}^* \varphi_{\bf k} \int \varphi_{\bf k}^{4} dt - \varphi_{\bf
k}^{2} \int \varphi_{\bf k}^{*} \varphi_{\bf k}^{3} dt \Biggr)
\Biggr] e^{i {\bf k} \cdot {\bf x}}, \label{two fn}
\end{eqnarray}
where $[\hat{O}, \hat{B}]_{(n)} = [\hat{O}, [\hat{O},
\hat{B}]_{(n-1)}]$ with $[\hat{O}, \hat{B}]_{0} = \hat{B}$. Note
that all the terms linear in $(\lambda \hbar)$ vanish and at order
$(\lambda \hbar)^2$ the mode-mixing terms contribute quadratic
terms of $\varphi_{\bf k}$ or $\varphi_{\bf k}^*$ to the scale
dependent part, which are negligible compared to the quartic terms
in Eq. (\ref{two fn}).

Finally, after inserting the solution (\ref{long sol}) into Eq. 
(\ref{two fn}) and
explicitly doing the integral, we obtain the two-point correlation
function of the form
\begin{equation}
G_0 (0, t)
\frac{\sin\Bigl(\sqrt{\frac{m_f}{2t}}r\Bigr)}{\sqrt{\frac{m_f}{2t}r}}
\exp \Bigl[-\frac{m_f r^2}{8 t } \Bigr] + \frac{(\lambda
\hbar)^2}{2} G_1 (0, t)\frac{\sin\Bigl(\sqrt{\frac{m_f}{2 \cdot 3
t}}r\Bigr)}{\sqrt{\frac{m_f}{2 \cdot 3t}r}} \exp \Bigl[-\frac{m_f
r^2}{8 \cdot 3 t } \Bigr],  \label{app 2-fn}
\end{equation}
where $G (0, t)$'s are slowly varying functions of $t$. The
leading and sub-leading terms of the second integral of Eq.
(\ref{two fn}) cancel each other and the third-leading term is the
second term of Eq. (\ref{app 2-fn}). The first and the second
terms of Eq. (\ref{app 2-fn}) are the Gaussian and the
non-Gaussian contribution, respectively,  to the correlation
function. Each term oscillates and is bounded by an envelope
determined by exponentially decreasing functions. The first term,
the Gaussian term, has a larger initial amplitude but decays more
rapidly than the second term, the non-Gaussian term. Therefore,
immediately after the phase transition, the first term determines
the domain size, which obeys the well-known Cahn-Allen scaling
relation \cite{boyanovsky,rivers,bowick,calzetta,kim-lee}
\begin{equation}
\xi_{\rm G} (t) = \sqrt{\frac{8 t}{m_f}}. 
\end{equation}
However, as the phase transition continues, the second term begins
to dominate over the first term and leads to the non-Gaussian
scaling relation
\begin{equation}
\xi_{\rm NG} (t) = \sqrt{\frac{8 \cdot 3 t}{m_f}} = \sqrt{3}
\xi_{\rm G}. 
\end{equation}

A few comments are in order. First, as the time for spinodal
instability goes on, higher order non-Gaussian terms begin to grow
comparable to and eventually become larger than the Gaussian and
lower order terms. A series of transitions of scale relation may
be expected from the Gaussian one to the non-Gaussian ones of
higher orders. In fact it cannot happen in a realistic scenario
because of the sampling of the mean field around the true vacuum
and the de-excitation of excited states at order $(\lambda
\hbar)^2$ or higher. The sampling of the mean field $\varphi_{\bf
k} (t)$ around the true vacuum after crossing the spinodal line
stops exponential growth and limits the duration of spinodal
instability. Also there will be equally excitation and
de-exitation, thus making virtually impossible for a system to
achieve a state with inverted population, the very high excitation
having the largest population. In reality there will be an
increased occupation of excited states near the Gaussian state and
the very highly excited states will have a small occupation
probability.  Second, though we focus on the real scalar field,
the result above can be readily extended to a scalar field with
$O(N)$ symmetry before the phase transition. The two-point
function of a complex scalar field is essentially the same as that
of the real scalar field \cite{kim-lee}. We thus anticipate that
other topological defects, such as monopoles, strings, and
vortices, formed from the scalar field with the appropriate
internal $O(N)$ symmetry may have the same correlation length as
domains. The non-Gaussian effects thus may decrease the defect
density.

Finally we conclude with the physical implication of non-Gaussian
effects on domain growth, DCC and the density of topological
defects. The RHIC at Brookhaven and the LHC at CERN in the near
future will produce heavy ion beams at very high energies that
will lead to deconfinement of quarks present inside individual
nucleons. It is anticipated that a quark-gluon plasma will be
formed and approach equilibrium over a time $\approx 10^{-23} s$
and then will evolve rapidly far from equilibrium through a chiral
phase transition. It has been suggested that the chiral phase
transition will be accompanied by the formation of DCC
\cite{wilczek}. From a theoretical point of view it was argued
that the linear $\sigma$ model may be a reasonable model, that
includes some of the symmetry properties such as chiral symmetry
of QCD, to establish the extent of DCC. Rajagopal and Wilczek have
advocated that the exponential growth of soft modes of the linear
$\sigma$ model may lead to a large correlation and hence a large
domain of condensate \cite{wilczek}. This mechanism has
similarities to the spinodal decomposition in QFT. The numerical
simulation based on the Hartree-Fock approximation, however,
results in smaller (1 $\sim$ 2 fm) DCC regions than the
anticipated ($>$ 10 fm) DCC regions \cite{boyanovsky2}.

On the other hand, we find that the non-Gaussian effects increase
the domain volume by an order of magnitude. This implies that a
chiral phase transition may result in the formation of a large DCC
domain, the decay of which eventually leads to a profuse
production of coherent pions. However any extraordinary
observation should be investigated carefully to find if there is a
linkage to the DCC. The present study of a real scalar field
serves only as a simple model that suggests the role of
non-Gaussian effects on domains, and implies that the density of
topological defects may be small. A realistic study of DCC
formation during the chiral phase transition would require working
with the linear $\sigma$ model. It may be anticipated that long
wavelength modes will play a similar role to enhance not only the
size of domains in quark-gluon plasma, condensed matter and the
early Universe but also the density perturbation from the second
order phase transitions in inflation scenarios. These studies are
in progress.

\acknowledgments

We would like to thank C. H. Lee and S. Sengupta for useful
discussions. This work was supported in part by the Natural
Sciences and Engineering Research Council of Canada. The work of
S.P.K was supported in part by the Korea Research Foundation
under Grant No. 2000-015-DP0080.

\end{document}